\documentclass[english,aps,preprint,showpacs]{revtex4}
\usepackage[T1]{fontenc}
\usepackage[latin1]{inputenc}
\usepackage{graphicx}

\makeatletter

\makeatletter

\makeatletter

\makeatletter

\makeatletter

\makeatletter

\makeatletter

\makeatletter

\makeatletter

\makeatletter

\makeatletter



\makeatother

\makeatother

\makeatother

\makeatother

\makeatother

\makeatother

\makeatother

\makeatother

\makeatother

\makeatother

\usepackage{babel}
\makeatother
\begin{document}

\title{Marangoni Convection in Binary Mixtures}

\author{Jie Zhang }

\email{jz26@phy.duke.edu}

\author{Robert P. Behringer}

\email{bob@phy.duke.edu}

\affiliation{Department of Physics and Center for Nonlinear and Complex Systems,
Duke University, NC 27708, USA}

\author{Alexander Oron}

\email{meroron@tx.technion.ac.il}

\affiliation{Department of Mechanical Engineering, Technion-Israel Institute of
Technology, Haifa 32000, Israel}

\date{\today}

\begin{abstract}
Marangoni instabilities in binary mixtures are different from those
in pure liquids. In contrast to a large amount of experimental work
on Marangoni convection in pure liquids, such experiments in binary
mixtures are not available in the literature, to our knowledge. Using
binary mixtures of sodium chloride/water, we have systematically investigated
the pattern formation for a set of substrate temperatures and solute
concentrations in an open system. The flow patterns evolve with time,
driven by surface-tension fluctuations due to evaporation and the
Soret effect, while the air-liquid interface does not deform. A shadowgraph
method is used to follow the pattern formation in time. The patterns
are mainly composed of polygons and rolls. The mean pattern size first
decreases slightly, and then gradually increases during the evolution.
Evaporation affects the pattern formation mainly at the early stage
and the local evaporation rate tends to become spatially uniform at
the film surface. The Soret effect becomes important at the later
stage and affects the mixture for a large mean solute concentration
where the Soret number is significantly above zero. The strength of
convection increases with the initial solute concentration and the
substrate temperature. Our findings differ from the theoretical predictions
in which evaporation is neglected. 
\end{abstract}


\maketitle

\section{INTRODUCTION}

When a horizontal pure-liquid thin film is sandwiched between a warm
solid at the bottom and a cool gas at the top, heat transfer can be
used to drive fluid flows. At the liquid-gas interface, the surface
tension is a function of the interfacial temperature, and usually
decreases with temperature. A small temperature perturbation along
the film surface may create surface-tension inhomogeneities which
in turn trigger fluid instabilities. Flows observed in such films
are known as B\'{e}nard-Marangoni convection or surface-tension-driven
convection. Several recent theoretical and experimental studies are
relevant here\cite{Schatz_prl95,swinney_jfm,Davis_arfm1987,Colinet_2000}.
Differing from thin films of pure liquids, the instabilities in binary-mixture
films occur because the surface tension is a function of both temperature
and solute concentration \cite{Bhattacharjee_pre,Joo_jfm,Oron_pre04,Podolny_pof,Podolny_pof06}.
Given a vertical temperature gradient across the film, the concentration
gradient may then be imposed by an independent source or generated
spontaneously by the Soret effect \cite{Platten_EurPJE04}. If the
thermal Marangoni and solutal Marangoni effects enhance each other,
a long-wavelength monotonic instability is possible, while if they
compete with each other, an oscillatory instability may occur instead
\cite{Bhattacharjee_pre,Joo_jfm,Oron_pre04,Podolny_pof,Podolny_pof06}.
In the past decade, theoretical work has led to predictions for the
linear and nonlinear stages of instabilities in thin films of binary
mixtures \cite{Bhattacharjee_pre,Joo_jfm,Oron_pre04,Podolny_pof,Podolny_pof06}.
By contrast, experiments with surface-tension-driven instabilities
in binary mixtures are not available in the literature, to our knowledge.
The lack of such experiments has motivated us to investigate pattern
formation for binary mixtures in thin films.\newline

\section{GOVERNING EQUATIONS AND PARAMETERS OF THE SYSTEM}

The governing equations of the system, including the Soret effect
are: \cite{Oron_pre04,Podolny_pof06} \begin{equation}
\nabla\cdot\vec{v}=0,\label{incompress}\end{equation}
 \begin{equation}
\partial_{t}\vec{v}+(\vec{v}\cdot\nabla)\vec{v}=-\rho^{-1}\nabla p+\nu\nabla^{2}\vec{v}-\vec{g},\label{velocity}\end{equation}
 \begin{equation}
\partial_{t}T+(\vec{v}\cdot\nabla)T=\kappa\nabla^{2}T,\label{temperature}\end{equation}
 \begin{equation}
\partial_{t}c+(\vec{v}\cdot\nabla)c=D(\nabla^{2}c+\alpha\nabla^{2}T).\label{concentration}\end{equation}
 Here $t$ is time, $\vec{v}$, $p$, $\rho$, $\nu$ and $\vec{g}$
denote the fluid velocity field, pressure, density, kinematic viscosity,
and gravitational acceleration, respectively, $T$ and $\kappa$ represent
the temperature and thermal diffusivity of the fluid, respectively,
$c$ is the concentration of the solute, $D$ is mass diffusivity
of the solute, and $\alpha$ is the Soret coefficient.

In Eq. (\ref{temperature}), the heat flux induced by the concentration
gradient, known as the Dufour effect, is usually very weak in liquids,
and thus neglected. At the solid-liquid interface, $z=0$, $\vec{v}$
satisfies the no-slip, no-penetration condition $\vec{v}=0$, the
temperature is fixed $T=T_{w}$, and the mass flux vanishes, $c_{z}+\alpha T_{z}=0$.
Here, the wafer temperature $T_{w}$ is constant. At a nondeformable
liquid-air interface, $z=h(t)$, the boundary conditions are given
by \cite{Oron_RevMP}. The heat and mass flux balances as well as
those for the normal and tangential stresses read, respectively, \begin{eqnarray}
k_{T}T_{z}+q(T-T_{a})+j{\cal L}=0,\label{Tsurface}\\
-\rho D(c_{z}+\alpha T_{z})+jc=0,\label{massimperm}\\
j=\rho(w-h_{t})=\hat{K}(T-T_{a}),\label{massflux}\\
-\frac{j^{2}}{\rho_{v}}=-p+2\mu w_{z}.\label{normbal}\\
\mu(\partial_{z}\vec{u}+\nabla w)=-\sigma_{T}\nabla T+\sigma_{c}\nabla c.\label{tangbal}\end{eqnarray}

Here, $k_{T}$ is the thermal conductivity of the liquid, $\vec{u}$
is the two-dimensional projection of $\vec{v}$ onto the $x-y$ (horizontal)
plane, $q$ is the heat transfer coefficient, assuming Newton's law
of cooling, $T_{a}$ is the constant room temperature, $j$ is the
evaporative mass flux, $\mu$ is the liquid viscosity, $\rho_{v}$
is the vapor density, and ${\cal L}$ is the latent heat of evaporation.
Also, we have assumed that the surface tension $\sigma$ is a linear
function of temperature $T$ and concentration $c$; hence, $\sigma(T,c)=\sigma_{0}-\sigma_{T}(T-T_{0})+\sigma_{c}(c-c_{0})$.
Here, $\sigma_{0}$ is the reference surface tension at $T=T_{0}$
and $c=c_{0}$. $\sigma_{T}\equiv-(\frac{\partial\sigma}{\partial T})_{T_{0}}$
and $\sigma_{c}\equiv(\frac{\partial\sigma}{\partial c})_{c_{c}}$.
$\Delta T$ is the temperature difference across the film.

The dimensionless parameters of the problem are \begin{eqnarray}
P=\frac{\nu}{\kappa},L=\frac{\kappa}{D},B=\frac{qh}{k_{T}},M=\frac{\sigma_{T}\Delta Th}{\rho\nu\kappa},\\
E=\frac{k_{T}\Delta T}{\rho\nu{\cal L}},K=\frac{k_{T}\hat{K}}{h{\cal L}},\chi=\frac{\alpha\sigma_{c}}{\sigma_{T}},\end{eqnarray}

respectively, the Prandtl, Lewis, Biot, Marangoni, evaporation, interfacial
resistance and Soret numbers. The estimated values of these dimensionless
parameters at the solute concentration of $2M/l$ (Mole per liter
of water) and at the temperature $30^{o}C$ are $P=7$, $L^{-1}=88$,
$\chi=0.04$, $E=2.\times10^{-4}$, $K=3.\times10^{-3}$, $B=6.\times10^{-3}$.
The Marangoni number depends on the instantaneous value of the film
thickness $h$ and in our experiments varies between $2000$ and $300$.
The estimate of the Soret number is based on the experimental data
from Ref. \cite{Gaeta_JPC82}.

\section{EXPERIMENTAL SYSTEM}

Before carrying out an experiment, the first important issue to be
resolved is the selection of a resonably non-volatile working liquid,
i.e. a binary liquid. We explored a wide range of candidate materials.
We first tried a mixture of ethanol/water and then abandoned it due
to the fast evaporation of ethanol. We also tested different combinations
of silicone oils with noticeable differences in surface tension between
them to avoid non-miscibility. Only a few silicone oils satisfying
the requirements of non-volatility and miscibility are commercially
available and are affordable. However, for all the possible pairs
of oils, either one or both are highly volatile at room temperature
or they are highly viscous at low temperature and volatile when heated.
The volatilities are typically much larger than that of pure water.
As a consequence, we did not choose binary mixtures of silicone oils
for our experiments. After much trial and error, we chose solutions
of sodium chloride (NaCl) with pure water as the working binary mixtures.
There is an additional advantage in choosing the mixture of NaCl/water
as a working liquid because its many physicochemical properties are
available in the literature, including surface tension as a function
of both concentration and temperature \cite{Matubayasi_JCIS99}, and
the variation of the Soret coefficient with temperature and concentration\cite{Gaeta_JPC82}
among others. The evaporation of water is significantly weaker than
that of other strongly volatile liquids, so that there is a reasonably
long time to take measurements before the film dries out completely.
Finally, the convective patterns are readily visible a short time
after a thin film is drawn. \newline

\begin{figure}
\includegraphics[width=0.8\columnwidth,keepaspectratio]{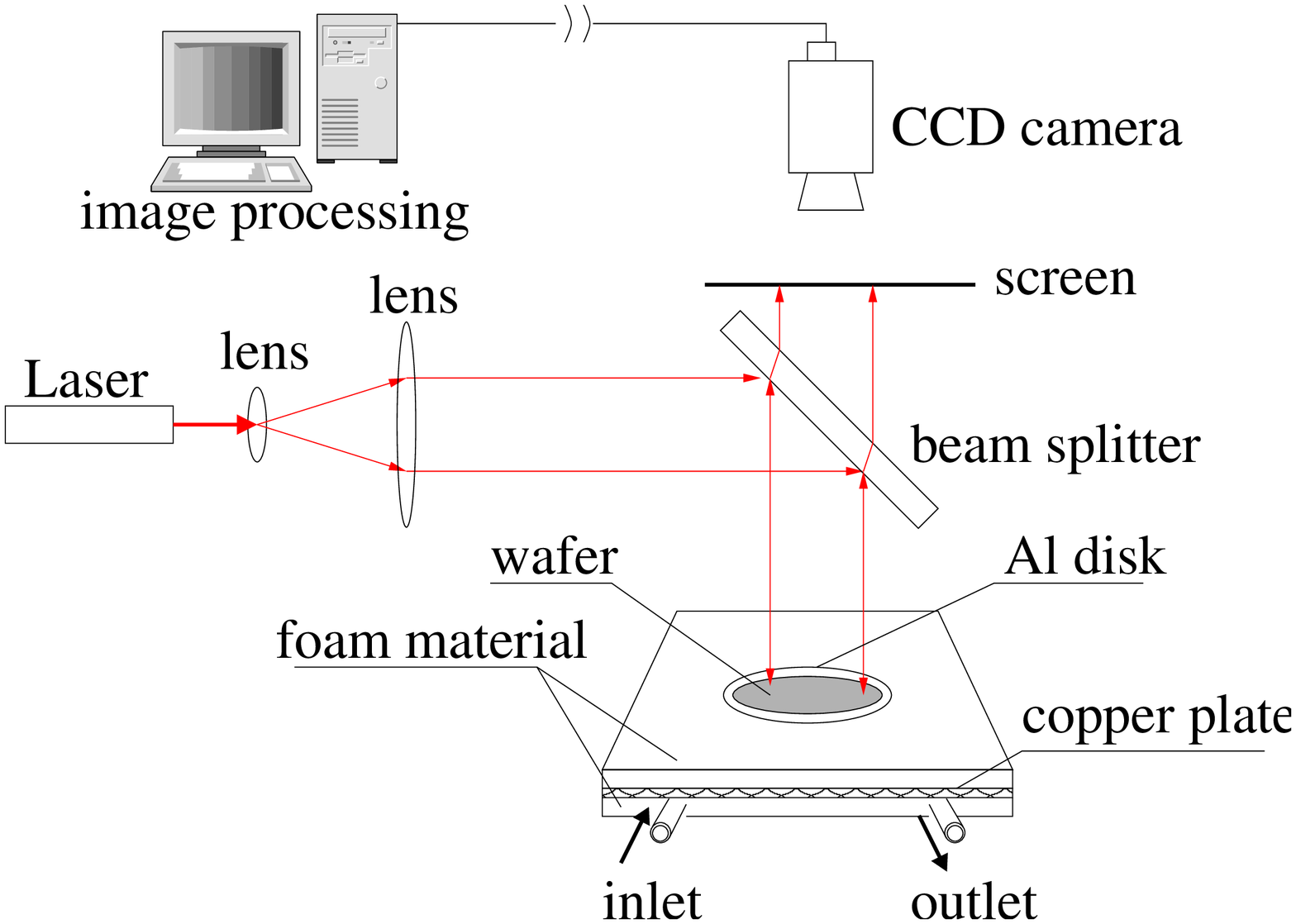}

\caption{\label{cap:setup}Schematic of the experimental setup. Visualization
is by standard shadowgraph using an expanded laser beam for a light
source. The bottom plate of the cell that contains the film consists
of a highly reflecting silicon wafer that is thermally grounded to
an highly conducting copper plate. This plate is temperature controlled
by means of a regulated water bath.}
\end{figure}

Figure \ref{cap:setup} presents a sketch of the experimental setup
of the system. The film was prepared on top of a silicon wafer with
a circular Plexiglas ring (about $7\, cm$ in diameter) as the lateral
boundary. The wafer was bonded to an Aluminum disk attached to a thin
copper plate. A copper tubing coil was soldered to the bottom of the
latter. The two ends of the coil were connected to a water bath (Neslab
RTE-221) using thermally insulated plastic tubing. The temperature
of the plate was held constant by water circulation. To further reduce
the heat loss, the whole copper plate was covered by thermally insulating
foam materials, leaving only a circular hole to hold the Aluminum
disk. The film was directly exposed to the room air whose temperature
is approximately constant during the course of an experiment. The
fluctuation in the room air is $\pm0.5\,^{o}C$, equivalent to $\sim0.5\times B\,^{o}C\approx3\, mK$
of the film temperature stability. The temperature fluctuation of
the wafer was affected by several sources. The first source is the
thermal fluctuation of the Neslab water bath which does not exceed
$10\, mK$. The second source is the temperature fluctuation at the
liquid-air interface. To further reduce the thermal fluctuation at
the silicon wafer, a bridge-controlled heating method was applied\cite{beh-ahlers}.
The wafer temperature was measured by a thermistor embedded underneath
which was measured with an AC bridge. An offset of the bridge balance
point drove an integro-differential feedback circuit which, in turn,
drove a heater attached to the inlet of the tubing that carried the
cooling water. In such a way, the thermal fluctuation of the wafer
was further reduced. Depending on the mean temperature of the film,
the long-time thermal fluctuations of the wafer temperature are less
than $30\, mK$ at the mean temperature of the film $T_{f}\equiv T_{w}-\frac{1}{2}\bigtriangleup T\approx30\,^{o}C$
and less than $15\, mK$ at $T_{f}\approx25\,^{o}C$. The mean temperature
drop across the film $\bigtriangleup T$ is typically less than $1\,^{o}C$.
\newline

\section{RESULTS AND DISCUSSION}

\begin{figure}
\includegraphics[width=0.8\columnwidth,keepaspectratio]{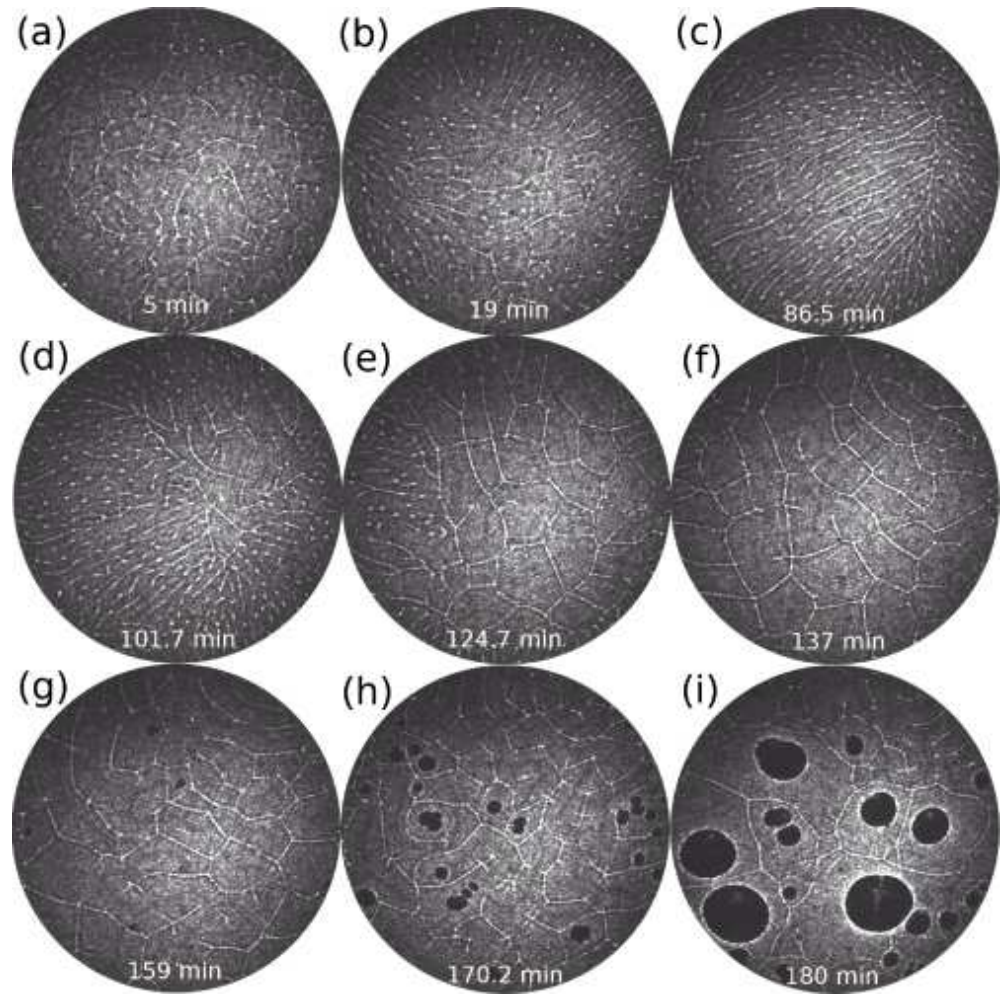}

\caption{\label{cap:pattern formation} Convection patterns observed at a
film temperature $T_{f}=30\,^{o}C$ and a room temperature $T_{a}=20\,^{o}C$.
The initial concentration of NaCl is $c_{0}=4\, M/l$ and the initial
film thickness is $h_{0}=1.8\, mm$. Images (a-i) show the flow pattern
at nine different times $t$. Bright and dark shades correspond to
downward and upward flows, respectively. The estimated initial Marangoni
number is $M_{0}=2122$. }
\end{figure}

Figure \ref{cap:pattern formation} shows a set of typical shadowgraph
flow patterns observed in thin films of NaCl/water mixtures with an
initial NaCl concentration of $4\, M/l$. In this figure, bright and
dark shades represent downward and upward flows, respectively. In
comparison with the standard Marangoni convection in pure liquids
(MCP), these patterns are very irregular and evolve with time. Following
the film deposition, an irregular pattern emerges, as seen in Fig.
\ref{cap:pattern formation}(a). This is a transient process with
patterns gradually decreasing in size (Fig. \ref{cap:pattern formation}(b)).
Eventually small-scale patterns cover the entire system, as shown
in Fig. \ref{cap:pattern formation}(c). The majority of the patterns
are roll-like structures, although there are also localized dots,
which are rarely observed in MCP. These small-scale patterns persist
for a period of time before any qualitative changes appear. Then,
the characteristic scale of the pattern becomes larger and the patterns
consist primarily of polygons, as shown in Fig. \ref{cap:pattern formation}(d-f).
If the concentration of NaCl exceeds the maximal solubility ($\approx8\, M/l$),
salt begins to crystallize out of the solution; small islands of salt
appear as black dots, as in Fig. \ref{cap:pattern formation}(g-i).
An interesting aspect of the system evolution concerns the film thickness.
Interferometric measurements show that except at the meniscus boundary,
the film thickness remains horizontally \textit{{uniform}} during
the evolution preceding the formation of salt crystals when deformation
of the film interface becomes significant.\newline

\begin{figure}
\includegraphics[width=0.8\columnwidth,keepaspectratio]{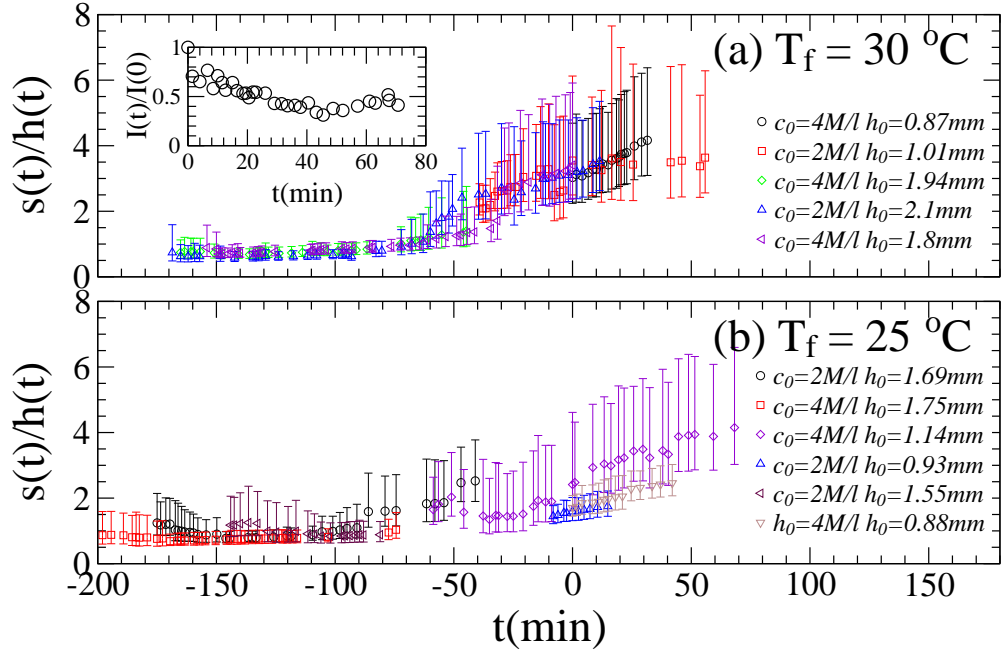}

\caption{\label{cap:evolution of pattern size} The average normalized pattern
size $s(t)/h(t)$ as a function of time $t$. The time axis is chosen
(thus negative times) so that different data sets have approximately
the same film thickness $h(t)=0.9\, mm$ at $t=0\, min$. Various
symbols represent measurements performed at different initial concentrations
$c_{0}$ of NaCl, different initial film thickness $h_{0}$, and different
film temperatures $T_{f}$. The inset shows the normalized pattern
intensity $I(t)/I(0)$ of the data set corresponding to the open circles. }
\end{figure}

To quantify the time variation of the pattern size, we apply fast
Fourier transforms (FFT) to the original shadowgraph images after
the removal of the background intensity variation. The images in the
Fourier space were averaged over phase angles. From the FFT spectrum,
we calculate both the mean wavenumber and its standard deviation.
The results are then converted to a real space pattern size, $s(t)$,
as presented in Fig. \ref{cap:evolution of pattern size}(a). We normalize
this characteristic pattern size by its instantaneous film thickness,
which we refer to hereafter as the normalized pattern size. Immediately
after the film is initialized, the normalized pattern size is approximately
$0.8$, i.e. the average pattern size is slightly below the value
of the film thickness. For some time thereafter, the patterns are
rolls, as seen in Fig. \ref{cap:pattern formation}(c). The normalized
pattern size also retains its initial value for a considerable time,
and then gradually increases. Due to evaporation, the film thickness
decreases monotonically. Because the Marangoni number is proportional
to the film thickness, it too continuously decreases. Finally, the
average normalized pattern size reaches $s/h\simeq3.0$. At this point,
the corresponding pattern consists mainly of non-equilateral polygons.
The irregular shape of these polygons leads to large errors in determining
the pattern size, Fig. \ref{cap:evolution of pattern size}. This
feature is different from the classical MCP where patterns are regular
polygons \cite{Benard,Schatz_prl95}. This difference may be caused
by one or a combination of several sources. First, there is the fact
that the surface conditions in the MCP are usually precisely controlled,
as compared to the present open system, for which a precise control
of the air layer above the film surface is much more difficult. Alternatively,
we note that the Marangoni number is ever changing, and it is impossible
to tune the value of the Marangoni number exactly at the threshold
value.\newline

Towards the later stages of the pattern evolution, a noticeable difference
in the pattern occurs as relatively linear rolls give way to polygonal
patterns. At that point, the Marangoni number is $539\pm29$, which
is quite big compared with the threshold values derived in non-evaporative
films \cite{Bhattacharjee_pre,Joo_jfm}, and the film thickness was
about $0.9$~mm. We also carried out an experiment with an initial
film thickness of $0.87\, mm$, which is close to the value mentioned
above. At this initial film thickness, the convection starts from
polygons and the pattern intensity decays with time until the convection
pattern vanishes. The magnitude of the pattern intensity, $I(t)$,
as a function of time is plotted in the inset of Fig. \ref{cap:evolution of pattern size}(a).
$I(t)$ is calculated from the average intensity of the image after
the removal of the background. A large Marangoni number was also observed
in an experiment using pure liquid films with strong evaporation \cite{Chai_ExpHT98}.
However, no patterns were observed in pure water films in our system.
This suggests the solutal effects (thermal solutal and Soret effect)
are important. This also leads us to the conjecture that a considerable
part of the temperature drop across the film may actually occur spatially
uniformly at the film surface. This is because evaporation can create
a large temperature drop across the film due to the latent heat loss.
The rest of the temperature drop across the film is responsible for
the fluid instabilities at the film surface. Indeed, we estimate that
as much as 95\% of the temperature drop across the fluid may occur
at the surface. Note that Fig. \ref{cap:evolution of pattern size}
does not show the data points where the crystallization of NaCl (black
dots) takes place at the film surface, as seen in Fig. \ref{cap:pattern formation}(g-i).
The existence of the solid phase changes the flow pattern dramatically.
Solid NaCl can grow and merge with a deformable fluid interface. The
physics of this stage has not yet been well understood and is outside
the scope of this work. \newline

Figure \ref{cap:evolution of pattern size}(b) shows the time evolution
of the normalized pattern size at a different temperature, $T_{f}=25\,^{o}C$.
While the qualitative features are similar to Fig. \ref{cap:evolution of pattern size}(a),
there are still some quantitative differences. The normalized pattern
size at the early stage is around $1.0$ which is larger than $0.8$
at $T_{f}=30\,^{o}C$. In the beginning of each run (see e.g. circles
and diamonds), the normalized pattern size usually decreases for the
initial $10\sim20\, min$. This feature exists in some of the other
runs in figure (a) or (b), except it is more visible here. We note
that typically, after the film is deposited, patterns generally appear
at a slightly bigger size; they then gradually decrease in size by
developing smaller scale convection polygons or rolls such as those
in Fig. \ref{cap:pattern formation}(a-c).\newline

\begin{figure}
\includegraphics[width=0.8\columnwidth,keepaspectratio]{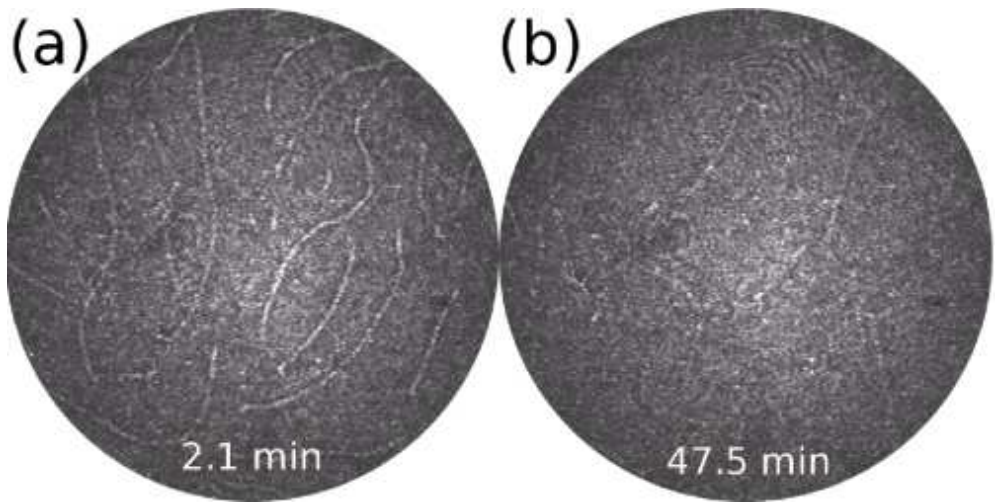}

\caption{\label{cap:patternweak} Weakening of the convection intensity at
$T_{f}\approx30\,^{o}C$, $T_{a}=22.2\,^{o}C$, initial concentration
of NaCl $c_{0}=0.4\, M/liter$, and initial film thickness $h_{0}=2.2\, mm$. }
\end{figure}

Figure \ref{cap:evolution of pattern size} does not present the normalized
pattern-size variation for films with low concentrations where the
convection flows are weaker and convection decays rapidly. For instance,
in Fig. \ref{cap:patternweak}(a) the patterns are readily seen at
time $t=2.1\, min$; they are very weak if not completely gone in
(b) at $t=47.5\, min$ after the film is deposited. Similar observations
were seen for initial concentrations $c_{0}\le0.8\, M/l$. The decay
of the pattern is an indication that the local evaporation flux has
a tendency to become spatially uniform with time. It is reasonable
to believe that the evaporation rate, and hence this effect, should
be independent of concentration. For higher concentrations, however,
the patterns persist and the intensity tends to be more uniform during
the evolution. One possibility is that this difference is related
to the Soret coefficient, as determined, for instance, from data for
NaCl/water mixtures by Gaeta et al.\cite{Gaeta_JPC82} At $T=30\,^{o}C$,
the Soret coefficient is $\alpha\simeq0$ at $c=0.2\, M/l$ and linearly
increases as $\log(c)$ to $\alpha=1.5\times10^{-3}/^{0}C$ at $c=2\, M/l$.
We estimate the corresponding Soret number to be $\chi=0.04$ for
$c=2\, M/l$ and $\chi=0.10$ for $c=4\, M/l$. The characteristic
time for mass diffusion, about $40\, min$ for a $2\, mm$ thick film,
is roughly the same as in Fig \ref{cap:patternweak}. This suggests
that the Soret effect becomes important at the later stage of convection.
In Fig. \ref{cap:patternweak}(b), the value $\chi\approx0.002$ is
too small to drive convection after the pattern created by evaporation
begins fading away. Thus, in order for the Soret effect to play a
role, a sufficiently large concentration leading to a higher Soret
diffusion is necessary since the solutal Marangoni effect is directly
proportional to the Soret number $\chi$.\newline

\begin{figure}
\includegraphics[width=0.8\columnwidth,keepaspectratio]{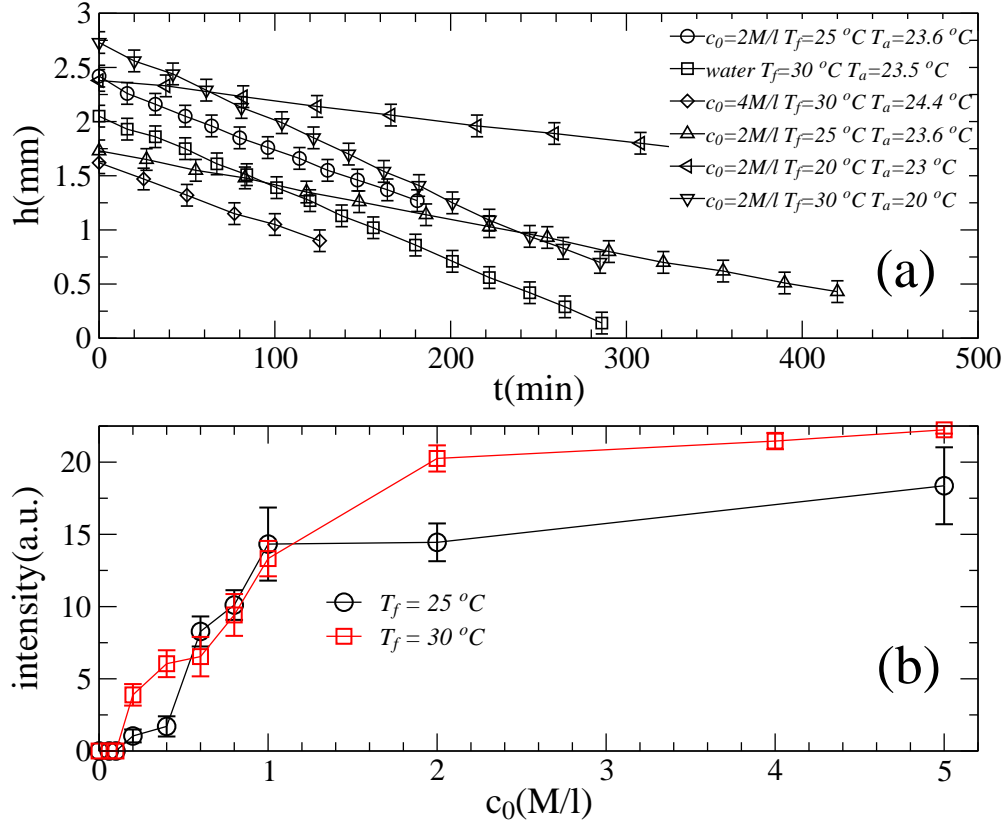}

\caption{\label{cap:brightness}(a) The film thickness change as a function
of time in the presence of evaporation. (b) The convection intensity
as a function of the initial NaCl concentration $c_{0}$ and film
temperature $T_{f}$ with the room temperature $T_{a}=20\,^{o}C$.
No convection is observed for $c_{0}<0.2\, M/l$. }
\end{figure}

In our experiments we have measured the instantaneous thickness of
the film over extended periods of time. Using the time history we
then evaluate the evaporation rate as a function of time. We find
that this quantity is nearly constant at a given film temperature
$T_{f}$ and a room temperature $T_{a}$. Considering that the mean
concentration increases with time, the linear plot for $h(t)$, as
seen in the Fig. \ref{cap:brightness}(a), implies a very weak concentration
dependence on the evaporation rate. The slope of $h(t)$ vs. $t$
increases for a larger film temperature $T_{f}$ or a smaller room
temperature $T_{a}$. The linear decrease of the film thickness due
to evaporation seems to be robust and almost independent of the NaCl
concentration. As a comparison, the measurements of the film thickness
for evaporating pure-water films are also presented in Fig. \ref{cap:brightness}(a).
The slope of $h(t)$ for a water film is also a constant, indicating
that this linear behavior is indeed independent of the composition
of the liquid film.\newline

This result is different from the analytical solution obtained by
Burelbach et al. \cite{Burelbach_jfm1988} and Oron et al.\cite{Oron_RevMP}
for the base state of an evaporating film of a uniform thickness \begin{equation}
h(t)=h_{0}[\sqrt{(K+1)^{2}-2Et}-K],\end{equation}
 where $h_{0}$ is the initial film thickness, $K$ is a constant
interfacial resistance to phase change and $E$ is the evaporation
constant. If $Et\ll1$, $h(t)$ would be nearly linear. However, as
$Et$ grows, $h(t)$ would start to bend downwards, especially at
small $h(t)$. This trend is not seen in the present data. We have
experimentally ruled out the possibility that convective air motion
above the film might be responsible, since the theory assumes that
the air layer is at rest. In particular, experimental measurements
of the evaporation rate for a film at a temperature lower than the
room air are also presented in the inset of Fig. \ref{cap:brightness}(a)
shown by the triangle-left symbols; they too show a linear variation.
We believe that the reason for the discrepancy between theory and
experiment is caused by the large fraction of the temperature difference
that occurs at the liquid-air interface. Because this temperature
difference remains nearly constant, the evaporation rate also remains
constant. Hence, the boundary condition used in recent theoretical
work differs from that which applies for these experiments. \newline

Keeping other conditions the same, the magnitude of the convection,
as determined based on the average intensity of the shadowgraph images
recorded over the first 15 minutes after the film was initialized,
depends on the initial concentration $c_{0}$ of NaCl and the film's
mean temperature. At low concentrations, $c_{0}\le0.1\, M/l$, there
are no visible convection patterns even for a $2.2\, mm$ thick film.
This implies that buoyancy is irrelevant in our system if we assume
that the thermal expansion of the fluid is not significantly affected
by adding NaCl to the pure water. When the concentration $c_{0}$
is higher than $0.2\, M/l$, the patterns gradually appear. This observation
suggests that thermal Marangoni alone cannot drive the fluid instability
and that the fluid motion is driven by the combination of thermal
Marangoni and solutal Marangoni effects. The magnitude increases monotonically
with concentration for $c_{0}\ge0.2\, M/l$. At a fixed film temperature
of $30\,^{o}C$, the Soret coefficient is negative in the concentration
range $0.042\, M/l\le c_{0}\le0.13\, M/l$\cite{Gaeta_JPC82}. However,
in such a range, the concentration is too low to observe any realistic
motions in the film. \newline

\section{SUMMARY}

To summarize, we have experimentally studied transient Marangoni convection
in thin binary-mixture liquid films using solutions of NaCl/water
in an open system. In the presence of evaporation, the patterns, consisting
of rolls and polygons, evolve with time. The length scale of the pattern
increases with time until the convection ceases. The Marangoni numbers
near onset are much larger than the values predicted in theories developed
for non-volatile binary liquids. Evaporation is important to the pattern
formation at the early stage, while the Soret effect is essential
at the later stage. For higher concentrations of NaCl, the convection
becomes stronger. Within the temperature range $25\,^{o}C\le T_{f}\le30\,^{o}C$
explored in the experiment, convection patterns were observed in the
NaCl concentration range $0.2\, M/l\le c_{0}\le5\, M/l$. Understanding
the rich pattern forming dynamics clearly requires new theoretical
work.\newline

\begin{acknowledgments}
This work was supported by NSF grant number $DMS-02444498$. J. Z.
thanks Matthias Sperl for his comments and suggestions of the manuscript.
J. Z. also enjoyed the discussion with Peidong Yu about the lockin
amplifier. A. O. acknowledges the hospitality of the focused research
group in thin films and of the Mathematics department of Duke University.
A. O. was partially supported by the Israel Science Foundation founded
by the Israel Science Foundation through Grant no. 31/03-15.3. \newline\newpage{}

\end{acknowledgments}
\bibliographystyle{h-physrev3.bst}
\bibliography{binaryconvection}

\end{document}